\documentclass[aps,prl,amsmath,amssymb,twocolumn,superscriptaddress]{revtex4}

\usepackage{amsmath}
\usepackage{color}
\usepackage{epsf}

\usepackage{graphicx}
\usepackage{psfrag}

\begin{document}


\title{Crystallization Mechanism of Hard Sphere Glasses}

\author{Eduardo Sanz}
\email{esanz@ph.ed.ac.uk}
\affiliation{SUPA, School of Physics and Astronomy, University of Edinburgh, Mayfield Road, Edinburgh, EH9 3JZ, Scotland}

\author{Chantal Valeriani}
\affiliation{SUPA, School of Physics and Astronomy, University of Edinburgh, Mayfield Road, Edinburgh, EH9 3JZ, Scotland}

\author{Emanuela Zaccarelli}
\affiliation{CNR-ISC and Dipartimento di Fisica, Universita' di Roma La Sapienza, P.le A. Moro 2, I-00185 Roma, Italy}

\author{W. C. K. Poon}
\affiliation{SUPA, School of Physics and Astronomy, University of Edinburgh, Mayfield Road, Edinburgh, EH9 3JZ, Scotland}

\author{P. N. Pusey}
\affiliation{SUPA, School of Physics and Astronomy, University of Edinburgh, Mayfield Road, Edinburgh, EH9 3JZ, Scotland}

\author{M. E. Cates}
\affiliation{SUPA, School of Physics and Astronomy, University of Edinburgh, Mayfield Road, Edinburgh, EH9 3JZ, Scotland}

\begin{abstract} 

In supercooled liquids, vitrification generally suppresses crystallization. Yet some glasses can still crystallize despite the arrest of diffusive motion. 
This ill-understood process may limit the stability of glasses, but its microscopic mechanism is not yet known.
Here we present extensive computer simulations addressing the crystallization of monodisperse hard-sphere glasses at constant volume (as in a colloid experiment). 
Multiple crystalline patches appear without particles having to diffuse more than one diameter. 
As these patches grow, the mobility in neighbouring areas is enhanced, creating dynamic heterogeneity 
with positive feedback. The future crystallization pattern cannot be predicted from the  
coordinates alone: crystallization proceeds by a sequence of stochastic micro-nucleation events, 
correlated in space by emergent dynamic heterogeneity.

\end{abstract}

\maketitle

A supercooled liquid is thermodynamically unstable towards crystallization, but may nonetheless survive for long periods.
For moderate supercooling, crystallization is delayed by a large free-energy barrier separating the metastable
fluid from the solid; thermal crossing of this barrier is followed by rapid deterministic growth of the resulting supercritical nucleus. This process is described fairly well by so-called `classical nucleation theory' (CNT) \cite{debenedetti_book}. For deep supercooling, the barrier is low and can be crossed by spatially diffuse collective motion sometimes compared to spinodal decomposition \cite{PhysRevLett.60.2665,cavagna,trudu}. In spinodal decomposition, as in CNT, the late stages involve deterministic nonlinear amplification of order parameter fluctuations \cite{Bray}. At very deep supercooling, motion can become so slow that the fluid's structure vitrifies, forming a glass, before crystallites can emerge. Though such structural arrest clearly inhibits crystal formation, it may not guarantee the glass to be long-lived. (Indeed, even glasses that appear permanent may unexpectedly crystallize in a process known as `devitrification' \cite{kelton}.) To reliably formulate stable glasses, a better mechanistic understanding is needed of how glasses can crystallize despite their structural arrest. 
  
It is challenging experimentally to study particle-scale dynamics in atomic or molecular systems,
due to the small size of their constituents. 
To overcome this problem, colloidal systems (whose particles are visible with microscopy) have been widely used to explore the physics of glasses \cite{S_2002_296_104,N_1992_362_616} and crystallization 
\cite{PhysRevLett.96.175701,S_2001_292_258,JCP_2009_130_084502}. 
Indeed, hard-sphere colloids (where density and supersaturation play the roles of inverse temperature and supercooling) are arguably the simplest system displaying fluid-solid coexistence and glass formation. They have therefore inspired many computer simulations   \cite{PRL_2003_90_085702,Nature_2001_409_1020,filion,PNAS_2010_107_14036,PhysRevLett.105.025701,JPCM_2010_22_232102}. Most of these address nucleation from the ergodic fluid; here CNT broadly holds although open questions remain \cite{Nature_2001_409_1020,filion,PNAS_2010_107_14036,PhysRevLett.105.025701}. Few simulation studies have addressed crystallization of glasses ---although this has been reported experimentally for colloids \cite{N_1992_362_616,JCP_2009_21_472118} (as well as molecules \cite{kelton})--- and no microscopic description of its mechanism yet exists. 

Here we explore the mechanism of glass crystallization (at fixed volume, as in colloidal experiments) at the level of individual crystallites and particles. We analyze the emergence of crystalline patches (of which none are initially present), finding that these appear and grow without needing {\em even a subset} of particles to move much beyond a diameter (a stronger result than in \cite{PRL_2009_103_135704}). The growth of crystalline patches enhances the mobility of the surrounding particles, creating an {\em emergent dynamic heterogeneity} (DH), not present initially, which gives positive feedback for further crystallization. Notably, even in the late stages, domain growth 
\textcolor{black}{is not controlled by barrier-free (quasi-deterministic \cite{Bray}) evolution of any configurational order parameter.}
Instead, the future pattern of crystallinity depends on thermally random particle {\em velocities}; reassigning these, even midway through the domain growth, leads to a different final pattern. 
This sensitivity resembles the stochastic, activated dynamics of early-stage CNT. Thus our crystallization mechanism for glasses can be viewed as a chaotic sequence of random micro-nucleation events, correlated in space by emergent DH. 
(To determine whether this process is related to the putative  `spinodal crystallization' of non-vitrified fluids at deep undercooling \cite{PhysRevLett.60.2665,cavagna,trudu,klein} requires detailed clarification of the latter mechanism.)

In the results presented here we use as units the particle mass $m$, 
diameter ${\sigma}$, and the thermal timescale $t_0=\sigma (m/k_B T)^{1/2}$ (so that $m=\sigma=k_BT=1$). 
Our event-driven molecular dynamics (MD) study addresses $N=86400$ monodisperse hard spheres in an $NVT$ ensemble, with periodic boundary conditions in a cubic box of volume $V$. \textcolor{black}{(For more details of the methods see \cite{rapaport,PRL_2009_103_135704,peterI}).}  
To analyze the results, each particle is first assigned a vectorial bond order parameter $d_6$~\cite{PhysRevB.28.784}.
Pairs of neighbours are then identified 
using the criterion of van Meel and Frenkel  \cite{Koos_PhD}, and a pair is deemed `connected' if the scalar product of their $d_6$ vectors exceeds $0.7$. (The prescription of \cite{Koos_PhD} avoids a dependence on an arbitrary cutoff distance used in earlier work \cite{peterI}). Each particle is then labeled as `crystalline' if it is connected with at least 6 neighbours, and as `amorphous' otherwise; the fraction of crystalline particles is then denoted $X(t)$. We further quantify the local degree of crystalline order around any particle through a scalar bond-order parameter $Q_6$ as defined in Ref. \cite{JCP_2008_129_114707}. 

In order to generate a glassy initial configuration without 
pre-existing crystallization nuclei, we compressed a
small system (400 particles) to a high packing fraction,
$\phi \approx 0.64$, and then replicate it periodically in space after 
checking that the fraction of solid-like particles is less than 0.005. 
The configuration thus generated was isotropically expanded to the desired density before starting the MD run. 
\textcolor{black}{In this way make sure there are no non-amorphous regions ($Q_6 > $ 0.25 \cite{JCP_2008_129_114707,JPCM_2010_22_232102}) 
in the initial configuration.}
Note that after a short transient, no discernible trace of the initial periodicity could be detected 
(Fig. \ref{colori}). \textcolor{black}{Periodic boundaries do not affect the appearance or the growth of crystallites 
provided that the system is large enough
to accommodate a good number of clusters (Fig. 1(b))}.

\begin{figure}[h!]
\begin{center}
\includegraphics[width=0.4\textwidth,clip]{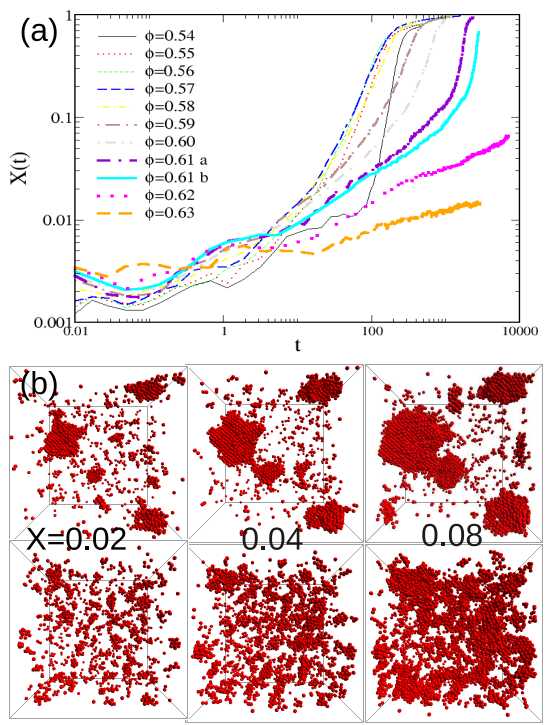}
\end{center}
\caption{\label{X} \textcolor{black}{(a) fraction of crystalline particles ($X$) versus time
for single runs at several packing fractions (see legend). The difference between curves 0.61 a and b is
discussed in Fig. 4.} 
(b) growth of the largest crystalline cluster for $\phi=0.54$ (top) and  $\phi=0.61$ (bottom). 
}
\end{figure}

Fig.~\ref{X}(a) shows the fraction of crystalline particles, $X(t)$, for various volume fractions $0.54 \le \phi \le 0.63$. 
At $\phi =  0.54$ the fluid is metastable for a short period  ($30 < t < 60$) until a post-critical cluster nucleates \textcolor{black}{and grows rapidly after $t\sim 100$}.  
For $\phi \ge 0.55$, $X(t)$ starts growing almost immediately ($t\sim t_0=1$): hence no significant nucleation barrier is present. 
Within the ergodic fluid at $0.55\le\phi\le 0.58$, $X(t)$ barely depends on $\phi$ \cite{peterII}, 
while in contrast, for glasses $(\phi > \phi_{g}\simeq 0.58)$, $X(t)$ grows more slowly initially, 
\textcolor{black}{but increases its slope sharply once sufficient particles have
crystallized to increase the free volume available to the remainder.}
Fig.~\ref{X}(b) contrasts the domain growth pattern of crystallites in the CNT regime ($\phi = 0.54$) with that in glass crystallization. 
Instead of a growing compact nucleus, 
a branched crystalline network develops. This is shown for $\phi=0.61$, the highest $\phi$ for which we observed crystallization within our time window (Fig.~\ref{X}(a)). From now on we focus on this case.

\begin{figure}[h!]
\begin{center}
\includegraphics[width=0.4\textwidth,clip=true]{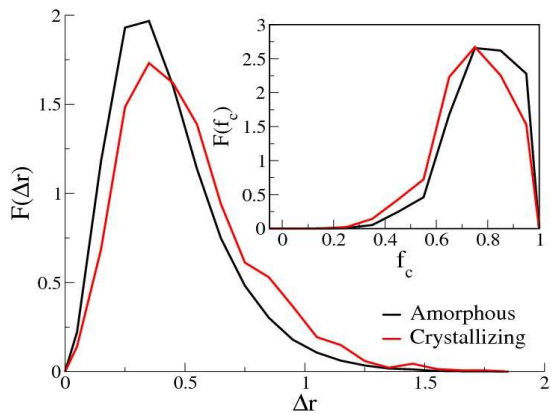}
\end{center}
\caption{\label{histo} 
Probability distributions of displacements between the start of the run and time $t_{0.1}$ defined by $X(t_{0.1}) = 0.1$. Two distributions are shown: in black are particles that were amorphous at $t=t_{0.08}$ and which remain amorphous at $t_{0.1}$. In red are particles that were amorphous $t=t_{0.08}$ but are crystalline at 
$t_{0.1}$. (Note that once a particle crystallizes it barely moves, which is why we examine only the mobility of particles that crystallized near the end of the selected time window at $t_{0.1}$.)
As can be seen in $F(\Delta r)$,  
most of the particles that crystallize do not move beyond one particle diameter during crystallization. Particles in the tail of the distribution do not have a crucial role as many crystalline regions appear without
any of these being involved.
Inset: probability distribution of the fraction $f_c$ of a particle's neighbours at time $t_{0.1}$ that were also neighbours at $t=0$ (`conserved neighbours'). Colour code is the same as in main figure. 
}
\end{figure}

In \cite{PRL_2009_103_135704} we established that  monodisperse hard sphere glasses can crystallize, even without pre-existing nuclei, while the root-mean-square (rms) particle displacement remains less than a particle diameter. Thus crystallization occurs without macroscopic diffusion --- which is why it remains possible within the glass. But that analysis did not rule out a form of DH involving larger diffusive displacements of the minority of particles actually involved in early-stage crystallization. This is however ruled out by Fig.~\ref{histo}, which shows separate displacement distributions for particles that become crystalline during a given time window and particles that remain amorphous: most crystallizing particles moved far less than a diameter since the quench. Fig.~\ref{histo} (inset) shows, for each population, the probability distribution $F$ with which a fraction $f_c$ of a particle's neighbours were also neighbours at the start of the run. Even for crystallizing particles, $\langle f_c\rangle \ge 0.7$. Thus crystallites indeed form without large diffusive rearrangements (cage-breaking), requiring only local shuffling of particles from amorphous into ordered patterns. Visual observation of particle trajectories confirms this \cite{supmov}.
Nonetheless, the mean-square displacements of crystallizing particles are measurably larger, 
and the mean $f_c$ smaller, than those of their amorphous counterparts (Fig.~\ref{histo}). 
This suggests a more subtle form of DH may contribute to the crystallization mechanism. 

\begin{figure}[h!]
\begin{center}
\includegraphics[width=0.4\textwidth,clip=true]{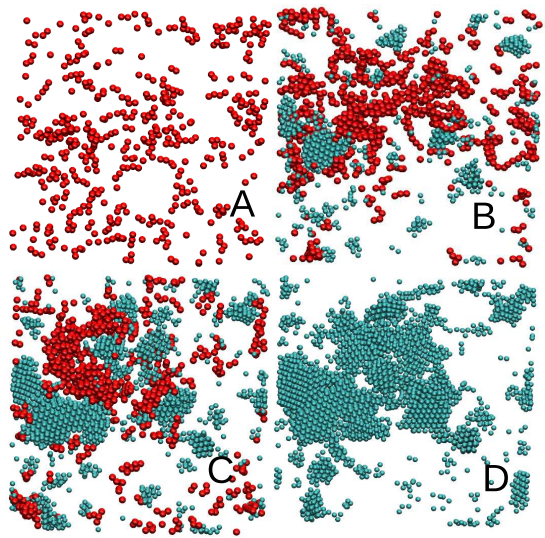}\\
\end{center}
\caption{\label{snapvelsol} 
Slab in the $xy$ plane showing the $5\%$ most mobile particles (in red) and the crystalline particles (in light blue) 
at time $t=0$ ($A$),  $t=320$ ($B$),  $t=640$ ($C$) and  $t=1280$ ($D$). Mobile particles are ranked by the distance they move between the 
time of the frame at which they are shown and the subsequent frame 
(accordingly, frame D does not show mobile particles).
Mobile particles are spatially correlated with crystalline ones and 
have a higher tendency to become crystalline than ``average'' amorphous particles.
}
\end{figure}

\begin{figure}[h!]
\includegraphics[width=0.4\textwidth,clip]{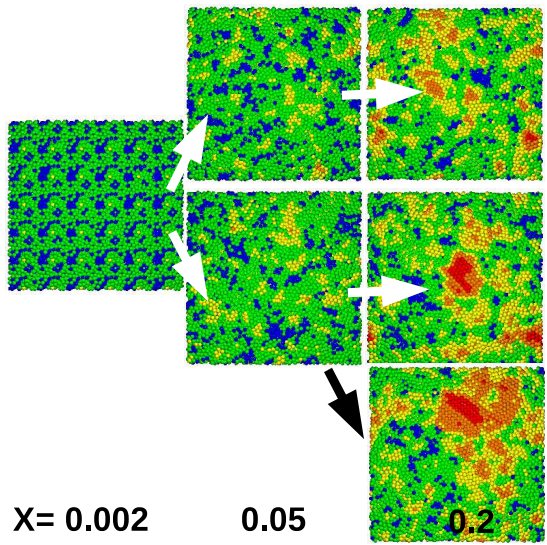}
\caption{\label{colori} Slab of the system for various values of the evolving crystallinity $X$. 
Particles are coloured according to the degree of crystalline order in their neighbourhood 
(blue, $0<Q_6<0.15$; green $0.15<Q_6<0.25$; yellow $0.25<Q_6<0.35$; orange $0.35<Q_6<0.45$; red $0.45<Q_6<0.55$). 
The initial state (left) has a periodic density pattern that is quickly forgotten. Its evolution follows two different trajectories 
(white arrows) from identical initial particle coordinates, but with different initial velocity choices (drawn at random from the thermal distribution).
These trajectories correspond to curves labelled as $\phi=0.61 a$ and $b$ in Fig. 1. 
The black arrow shows a run where velocities are randomized after $X$ reaches 0.05. The subsequent evolution is again altered, 
even though significant crystallinity was already present. At no stage do we find 
the future evolution to depend reproducibly on coordinates alone, although crystallites 
are more likely to form in regions of high $Q_6$ than elsewhere. 
}
\end{figure}

To elucidate its role, we present in Fig.~\ref{snapvelsol} maps showing the  
crystalline particles and the 5\% most mobile particles. 
While there is little DH initially, 
the most mobile particles become highly clustered once crystallites 
start to form, are preferentially 
found next to these crystallites,
and have a higher tendency to become crystalline thereafter.
The dynamics not only involves slow growth of crystallites but also the preferential 
creation of new ones next to the old, in regions of \textcolor{black}{enhanced 
mobility} (Fig.~\ref{snapvelsol} $B \rightarrow C$).
The crystallites are denser than their surroundings, as would be 
expected if pressures are fairly uniform (e.g., at $X=0.08$  crystallites have 
$\phi\simeq 0.65$ well above the global mean $\phi=0.61$). 
Thus our emergent DH is probably the result of an increase 
in free volume locally. Recently Ref. \cite{PNAS_2010_107_14036}
observed that particles at the interface between ordered and 
amorphous regions have lower density than average, although no feedback between DH and ordering emerged from that work.

Two runs starting from the same initial particle coordinates but with different (Maxwellian) initial velocities are shown in Fig.~\ref{colori}. The domain patterns (measured by $Q_6$) evolve quite differently. This confirms the absence of  pre-existing nuclei in our initial configuration: crystallinity arises at random locations. 
Indeed our initial coordinates, but not velocities, have a higher periodicity than the simulation box, which the emerging crystallinity pattern clearly ignores.
Moreover, even if the particle velocities are re-assigned midway through the domain growth, the subsequent evolution is markedly altered -- including that of crystallites that already exist (Fig. \ref{colori}). There are an infinite number of possible chaotic sequences of particle collisions in a given region; it appears that only some of these allow the crystalline order to increase. Thus the fate of a given region can be switched from 
amorphous to crystalline by changing either the initial or the mid-run velocities. \textcolor{black}{The stochastic nature of the 
process by which particles in the glass
incorporate to growing crystallites explains the difference between
curves 0.61 a and b in Fig. 1a.} As stated previously, this chaotic dynamics rules out any late-stage spinodal-like description 
\textcolor{black}{based on continuous free energy reduction by evolution of a configurational order parameter.}

Examples of such order parameters are not only the density but also orientational parameters, including $d_6$ and $Q_6$, as used to identify regions of 
`medium-range crystalline order' (MRCO). The role of such configurational order parameters in supercooled liquid dynamics is much debated \cite{cavagna}. For ergodic supercooled fluids, 
a two-step mechanism by which crystals nucleate within MRCO regions is increasingly accepted \cite{PhysRevLett.105.025701,PNAS_2010_107_14036,JPCM_2010_22_232102}. 
For crystallization from the glass, Fig.~\ref{colori}, high-$Q_6$ regions seen at late times likewise correspond to intermediate-$Q_6$ ones at earlier
stages (though in our simulations, none are present initially). However this conversion is quite gradual and does not involve an identifiable nucleation step. In \cite{JPCM_2010_22_232102} it was further argued that diffusionless crystallization might arise via ``positional
ordering in a region already having hcp-like bond orientational order''. Such regions need not be pre-existing, however: indeed all are eliminated from our initial conditions. 
A concept recently introduced to address DH in glasses is that of `propensity' \cite{propensity,NP_2009_4_711}. The basic idea of propensity is that configurational order does not directly determine the motion of a neighbourhood, but influences its probability of undergoing motion.  
\textcolor{black}{In this language an order parameter such as $Q_6$ could create a propensity for crystallization.}

Above we studied only `fresh' glasses, immediately post-quench, whose amorphous structure undergoes aging until crystallization intervenes. 
Using methods developed in \cite{peterII}, in future work we hope to find whether devitrification of aged glasses follows the same microscopic mechanism as described here for fresh ones.

In summary, we have elucidated the mechanism of crystallization of monodisperse hard-sphere glasses. 
As previously surmised \cite{PRL_2009_103_135704} amorphous regions gradually 
transform into crystallites \textcolor{black}{by `shuffling'  motion}. Crystalline regions grow in an autocatalytic way:
shuffles leading to crystallization create mobile regions within
which further crystallization is facilitated, giving an evolving dynamic heterogeneity that was not present initially. 
The local emergence of crystallinity is chaotic: its future course depends on the chance values of particle momenta, 
which determine which of many possible collisional trajectories actually arises. 
Structural order parameters do not fully determine where crystallites will form or how they will grow. 
The mechanism could be experimentally validated via particle tracking in colloidal suspensions
provided that the polidispersity allows for crystallization without fractionation
(up to 3-4\%).

{\bf Acknowledgements:} 
This work has made use of the Edinburgh Compute and 
Data Facility (ECDF), and was funded in part by EPSRC EP/E030173 and EP/D071070. 
MEC is funded by the Royal Society
and ES and CV by an Intra-European Marie Curie Fellowship. 
EZ, MEC and WCKP acknowledge support from ITN-234810-COMPLOIDS.

\bibliographystyle{./apsrev}

\end{document}